\documentclass{IOS-Book-Article}

\usepackage{mathptmx}
\usepackage{color}
\usepackage{tabularx}
\usepackage{amsmath}
\usepackage{listings}
\usepackage{titlesec}
\usepackage{epsf} 
\usepackage{rotating}
\usepackage{xcolor}
\usepackage{xspace}
\def\hb{\hbox to 10.7 cm{}}

\begin{document}

\pagestyle{headings}
\def\thepage{}

\newcommand{\hsml}{\texttt{hsml} }
\newcommand{\bchr}{\textbf{box5hr} }
\newcommand{\average}[1]{\langle #1\rangle}
\newcommand{\pd}{\code{particle\_data}}
\newcommand{\hpd}{\code{head\_particle\_data}}
\newcommand{\spd}{\code{sph\_particle\_data}}
\newcommand{\nsome}[1]{N_\text{#1}}
\newcommand{\ngroup}{\nsome{group}}
\newcommand{\nngb}{\nsome{ngb}}
\newcommand{\ncandidates }{\nsome{candidates}}
\newcommand{\magneticum}{\textit{Magneticum} }
\begin{frontmatter}              

\title{Exploiting the Space Filling Curve Ordering of Particles in the Neighbour Search of Gadget3}

\markboth{}{July 2015\hb}

\author[L,X,U]{\fnms{Antonio} \snm{Ragagnin}},
\author[T]{\fnms{Nikola} \snm{Tchipev}},
\author[T,L]{\fnms{Michael} \snm{Bader}},
\author[U,M]{\fnms{Klaus} \snm{Dolag}} and
\author[L]{\fnms{Nicolay} \snm{Hammer}}

\address[L]{Leibniz-Rechenzentrum (LRZ), Boltzmannstrasse 1, D-85748 Garching, Germany}
\address[X]{Excellence Cluster Universe, Boltzmannstrasse 2, D-85748 Garching, Germany}
\address[U]{Universit\"ats-Sternwarte, Fakult\"at f\"ur Physik, Ludwig-Maximilians Universit\"at M\"unchen, Scheinerstrasse 1, D-81679 M\"unchen, Germany}
\address[T]{Department of Informatics, Technical University Munich (TUM), Boltzmannstr. 3, D-85748 Garching, Germany}
\address[M]{Max-Planck-Institut f\"ur Astrophysik, Karl-Schwarzschild Strasse 1, D-85748 Garching bei M\"unchen, Germany}

\begin{abstract}

Gadget3 is nowadays one of the most frequently used high performing parallel codes for cosmological hydrodynamical simulations. Recent analyses have shown that the Neighbour Search process of Gadget3 is one of the most time-consuming parts. Thus, a considerable speedup can be expected from improvements of the underlying algorithms.

In this work we propose a novel approach for speeding up the Neighbour Search which takes advantage of the space-filling-curve particle ordering. Instead of performing Neighbour Search for all particles individually, nearby active particles can be grouped and one single Neighbour Search can be performed to obtain a common superset of neighbours.

Thus, with this approach we reduce the number of searches. On the other hand, tree walks are performed within a larger searching radius.
There is an optimal size of grouping that maximize the speedup, which we found by numerical experiments. 

We tested the algorithm within the boxes of the Magneticum   project. 
As a result we obtained a speedup of $1.65$ in the Density and of $1.30$ in the Hydrodynamics computation, respectively, and a total speedup of $1.34.$

\end{abstract}

\end{frontmatter}

Gadget3 (GAlaxies with Dark matter and Gas intEracT) simulates the evolution of interacting Dark Matter, gas and stars in cosmological volumes \cite{gadget2014,springel2005}. While Dark Matter is simulated so it interacts only through gravity, gas obeys the laws of hydrodynamics.
Both Dark Matter and gas are simulated by a particle approach.
Gadget3 uses a Tree-PM (see, e.g. \cite{xu1995parallel}) algorithm for the gravitational interactions between both Dark Matter and gas particles.
Smoothed Particle Hydrodynamics (SPH) is used for the hydrodynamic interaction, as described in \cite{beck2015improved}.

Gadget3 employs a variety of physical processes, e.g. gravitational interactions, density calculation, hydrodynamic forces, transport processes, sub-grid models for star formation and black hole evolution.
All these algorithms need to process a list of active particles and find the list of nearby particles (``neighbours'').
These neighbours are typically selected within a given searching sphere, defined by a given searching radius, defined by local conditions of the active particles (see, e.g. \cite{hernquist1989treesph}).
This problem is called Neighbour Search and is one of the most important algorithms to compute the physics implemented in Gadget3.

\section{Neighbour Search in Gadget3}

Simulations of gravitational or electromagnetic interactions deal with potentials having, ideally, an infinite range.
There are several known techniques (e.g. Barnes-Hut \cite{barnes-hut1986}, Fast Multipole Expansion\cite{greengard1997new}) that can deal with this problem.
These techniques subdivide the interaction in short-range and long-range interactions.
The Long-range interactions are resolved by subdividing the simulated volume in cubes, and assigning to each of them a multipole expansion of the potential.
The short-range potential is usually evaluated directly.
This leads to the problem of efficiently finding neighbours for a given target particle, within a given searching radius. 
Finding neighbours by looping over all particles in memory is only suitable when dealing with a limited number of particles.
Short-distance neighbour finding can be easily implemented by a Linked-Cell approach.
Since long-distance computation is implemented subdividing the volume in a tree (an octree if the space is three-dimensional), this tree structure is commonly used for short-distance computations too.
This is also a more generic approach, since Linked-Cell is more suitable for homogeneous particle distributions.

\subsection{Phases of Neighbour Search}

In Gadget3, the Neighbour Search is divided into two phases. The first phase searches for neighbours on the local MPI process and for boundary particles with possible neighbours of other MPI processes.
The second phase searches for neighbours in the current MPI process, of boundary particles coming from others MPI processes.
In more detail, the two phases of the Neighbour Search can be summarized in the following steps:
\begin{itemize}
\item First phase:
\begin{itemize}
	\item for each internal active particle $P_i$: walk the tree and find all neighbouring particles closer than the searching distance $h_i$;
	\item when walking the tree: for every node belonging to a different MPI process, particle and external node are added to an export buffer;
	\item if the export buffer is too small to fit a single particle and its external nodes: interrupt simulation.
	\item if the export buffer is full: end of first phase.
	\item physical quantities of $P_i$ are updated according to the list of neighbours obtained above.
\end{itemize}
\item Particles are exported.
\item Second phase:
\begin{itemize}
	\item for each guest particle $P_i$: walk the tree and search its neighbours;
	\item update properties of $P_i$ according to the neighbours list;
	\item send updated guest particles back to the original MPI process.
\end{itemize}
\item Current MPI process receives back the particles previously exported and updates the physical quantities merging all  results from the various MPI processes. 
\item Particles that have been updated are removed from the list of active particles.
\item If there are still active particles: start from the beginning.
\end{itemize}
The definition of neighbouring particles is slightly different between the Gadget3 modules.
In the Density module, neighbours of the particle $P_i$ are all the particles closer than its searching radius $h_i.$ 
In the Hydrodynamics module, neighbours are all particles $P_j$ closer than $max(h_i,h_j)$ to $P_i.$

\subsection{Impact of the Neighbour Search in the Gadget3 Performance}

\begin{figure*}

\begin{tabular}[t]{| l r  |}
\hline
 Hydrodynamics Routines & Time $[s]$ \\  
\hline
First Phase & $3.21\cdot10^{5}$   \\ 
\hline
First Phase Neighbour Search & $1.89\cdot10^{5}$  \\ 
\hline
Second Phase & $9.81\cdot10^{4}$  \\ 
\hline
First Phase Neighbour Search & $7.36\cdot10^{4}$  \\ 
\hline
\end{tabular}
\begin{tabular}[t]{| l r  |}
\hline
Summary Hydrodynamics & Time $[s]$ \\  
\hline
Physics & $1.55\cdot10^{5}$  \\ 
\hline
Neighbour Search & $2.63\cdot10^{5}$    \\ 
\hline
Communication & $7.17\cdot10^{4}$   \\ 
\hline
\end{tabular}
\caption{Left: \texttt{Scalasca} timing of the most expensive routines of the Hydrodynamics module in Gadget3. Right:  Aggregate timing of the Hydrodynamics parts.}
\label{table:skaf}
\end{figure*}

Tree algorithms are suitable for studying a wide range of astrophysical phenomena \cite{hernquist1987performace,warren1995portable}.
To perform the Neighbour Search in Gadget3, an octree is used to divide the three dimensional space.
Further optimization is obtained by ordering the particles according to a space-filling curve.
 In particular, Gadget3 uses the Hilbert space-filling curve to perform the domain decomposition and to distribute the work among the different processors.

We analysed the code with the profiling tool \texttt{Scalasca} \cite{geimer2010scalasca}.
In Figure  \ref{table:skaf} (left table) we show the profiling results for the Hydrodynamics module, which is the most expensive in terms of time.

The Hydrodynamics module is called once every time step.
It calls the First Phase and the Second Phase routines multiple times.
While the First Phase updates the physical properties of local particles, the Second Phase deals with external particles with neighbours in the current MPI process.
Particles are processed in chunks because of  limited exporting buffers, so the number of times those functions are called depends on the buffer and data sizes.
Between the First Phase and Second Phase calls there are MPI communications that send guest particles to others MPI processes.
First Phase and Second Phase routines are the most expensive calls inside Hydrodynamics, Density and Gravity.
Both the First Phase and the Second Phase perform a Neighbour Search for every active particle.

In Figure  \ref{table:skaf} (right table), the Hydrodynamics part has been splitted into three logical groups: Physics, Neighbour Search and Communication.
Communication between MPI processes has been estimated a posteriori as the difference between the time spent in Hydrodynamics  and the sum of the time spent in the First Phase and Second Phase.
This is well justified because no communications between MPI processes are implemented inside First Phase and Second Phase.
The time spent in Physics has been computed as the difference between the First (or Second) Phase and the Neighbour Search CPU time.
From this profiling, it turns out that for the Hydrodynamics module, Communication and Physics take less time than the Neighbour Search.
This was already suggested by a recent profiling \cite{karakasis2014}.
Both results highlight the interest  in speeding-up the Gadget3 Neighbour Search.

The three most time consuming modules in Gadget3 are Hydrodynamics, Density and Gravity.
In this work we only improved Density and Hydrodynamics modules.
There are two main reasons for excluding the Gravity module from this improvement.
First, Gravity module implements a Tree-PM algorithm \cite{bagla2002treepm}. 
Unlike in Density and Hydrodynamics, particles do not have a defined searching radius.
In fact the criterion whether or not a node of the tree  must be opened take into account the subtended angle of this node by the particle.
Also, for the way it is implemented in Gadget3, the Gravity module does not makes a clear distinction between the Neighbour Search and the physics computations, making it difficult to modify the Neighbour Search without a major rewriting of the module.

\section{Neighbour Recycling in Gadget3}

\begin{figure*}
\includegraphics[width=1.0\textwidth]{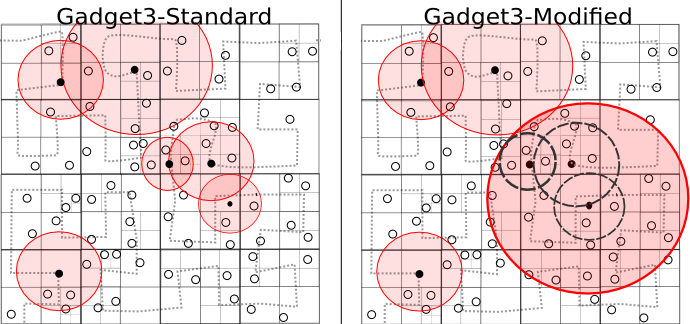}
		\caption{Difference between the standard method of Neighbour Search and the new one. Left panel contains the standard method, where for each active particle a new Neighbour Search is performed from scratch within a given searching radius $h_i$. Right panel contains the modified version, where particles are grouped within a certain radius $R$ and the Tree Walk is performed only once for a group of particles, within a searching radius of  $R+max(h_i)$.
In both panels, the octree is represented by square subdivisions, red circles represent the Neighbour Search radius, full dots represent active particles and empty dots represent inactive particles. The dashed line indicates the Hilbert curve.}
		\label{fig:beforeafter}
	\end{figure*}

We now show a novel approach to speed up the Neighbour Search.
It takes advantage of the space-filling-curve particle ordering in Gadget3.
As the locality of particles in memory maps to the locality of particles in the three dimensional space, consecutive  active particles share a significant part of their neighbours.
Therefore, nearby active particles are grouped and one single Tree Walk is performed to obtain a common superset of neighbours.
By that we reduce the number of tree walks.
On the other hand, tree walks are performed within a larger searching radius.
A sketch of the algorithm change is shown in Figure \ref{fig:beforeafter}.

In addition, the speedup gained by reducing the number of tree walks is lowered by the extra work to filter the true neighbours of each active particle from the superset of neighbours.
Thus, we may expect that there is an optimal grouping size to maximize the speedup, which can be determined by numerical experiments.

A  common Molecular Dynamics technique to recycle neighbours is the Verlet-List algorithm \cite{verlet1967}.
In the Verlet-List approach, a superset of neighbours is associated to each particle which is used within multiple time steps.
In our approach we associate a superset of neighbours to multiple particles, within a single time step.
This technique takes into account that two target particles which are close together will also share part of the neighbours.

Neighbour Recycling groups can be built by using the underlying octree structure.
Each  group can be defined as the set of leaves inside  nodes which are of a certain tree depth.
Then, a superset of neighbours is built by searching all particles close to that mentioned node. 
For each previously grouped particle, this superset of neighbours is finally refined.
An advantage of this technique is that the number of tree walks is reduced, though at the expense of a larger searching radius.

The level of the tree at which the algorithm will group particles determines both the number of tree walks and the size of the superset of neighbours.
This superset must be refined to find the true neighbours of a particle.
Thus, increasing its size will lead to a more expensive refinement.

\subsection{Implementation of Neighbour Recycling using a Space Filling Curve}

Many parallel codes for numerical simulations order their particles by a space-filling curve, mainly because this supports the domain decomposition  \cite{bungartz2010pde,gibbon2005performance,liu2000experiences}.
In this work we will benefit from the presence of a space-filling curve to implement a Neighbour Recycling algorithm. 
Due to the nature of space-filling curves, particles processed consecutively are also close by in space.
Those particles will then share neighbours.

Given a simulation for $N$ particles, our algorithm proceeds as follows.
A new group of particles is created, and the first particle is inserted into it.
A while loop over the remaining particles is executed.
As long as these particles are closer than a given distance $R$ to the first particle of the group, they are added to the same set of grouped particles.
Once a particle is found, which is farther than $R,$ the group is closed and a new group is created with this particle as first element.
The loop above mentioned is repeated until there are no more particles.
We call $\ngroup $ the number of particles in a given group; $h_i$ the searching radius of the $i$-th particle of the group.
Then, a superset of neighbours is obtained by searching the neighbours of the first particle of the group, within a radius of $R+max(h_i).$
This radius ensure that all neighbours of all grouped particles are included in the searching sphere.
For each grouped particle, the superset of neighbours is refined to its real list of neighbours.
The refined list of neighbours is then used to compute the actual physics quantities of the target particle.
Thus, the number of tree walks is reduced by a factor equal to the average number of particles in a group, $\average{\ngroup }.$

It is clear that a too low value of $R$ will group too few particles and lead to $\average{\ngroup }\simeq1$, thus leading to no noticeable performance increase.
On the other hand, if $R$  is too large, the superset of neighbours will be too large with respect to the real number of neighbours, producing too much overhead.

\subsection{Adaptive Neighbour Recycling}

In typical cosmological hydrodynamical simulations performed by Gadget3, a fixed $R$ will group less particles in low-density regions and more particles in high density regions.
Therefore a more reasonable approach is to reset $R$ before every Neighbour Search and choose it  as a fraction of the searching radius $h_i$, which itself is proportional to the local density.
In this way, low density regions will have a larger $R$ than high density regions. 
This is obtained by imposing the following relation between $R$ and the searching radius $h$ of the grouped particles:
$$R=f\cdot h_0,$$
where $f$ is a constant defined at the beginning of the simulation.

In a typical Gadget3 simulation, the number of particles $\nngb $ within the searching radius $h_i$ is a fixed quantity.
Locally it varies only within a few percent.
In the approximation that every particle has the same number of neighbours $\nngb $, we can write it as $\nngb =4\pi\rho{h^3_i}/3,$ where $\rho$ is the local number density.
Furthermore, if the grouping radius is small enough, the density does not vary too much and we can set $h_i=h.$
With those two approximations, the superset of neighbours is $\ncandidates =4\pi\rho(R+h)^3/3$ and the number of particles in a group is $\ngroup =4\pi\rho R^3/3.$
Combining those relations we obtain the following relation:
\begin{equation}
\label{eq:f}
f=\left(\frac{\ncandidates }{\nngb }\right)^{\frac{1}{3}}-1 = \left(\frac{\ngroup }{\nngb }\right)^{\frac{1}{3}}
\end{equation}

\subsection{Side Effects of Neighbour Recycling}

The Neighbour Recycling algorithm will increase the communication.
Because tree walks are performed within a larger radius, the number of opened nodes increases.
As a direct consequence, nodes of the tree belonging to other MPI processes will be opened more times than the original version.
In the standard approach, the export buffer is filled only with particles whose searching sphere intersect that node.
Since the new approach walks the tree for a group of particles, all particles belonging to the group are added to the export buffer.
This leads to a greater amount of communications.

\section{ Speedup of the Recycling Neighbours Approach on Gadget3}

\begin{figure}
\centerline{
\includegraphics[width=0.5\textwidth]{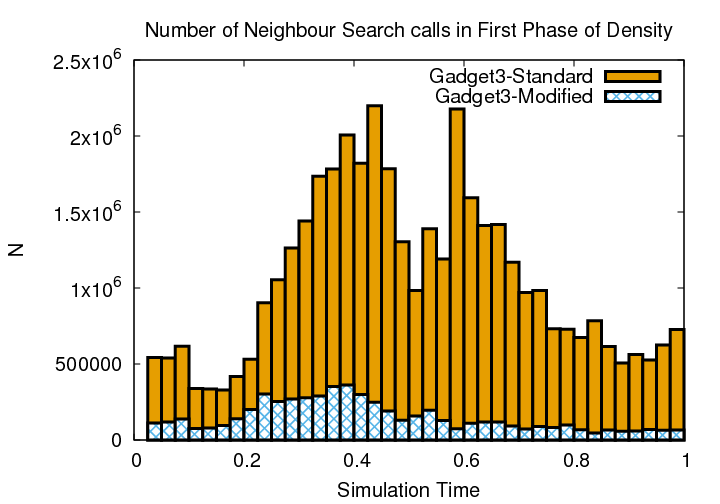}
\includegraphics[width=0.5\textwidth]{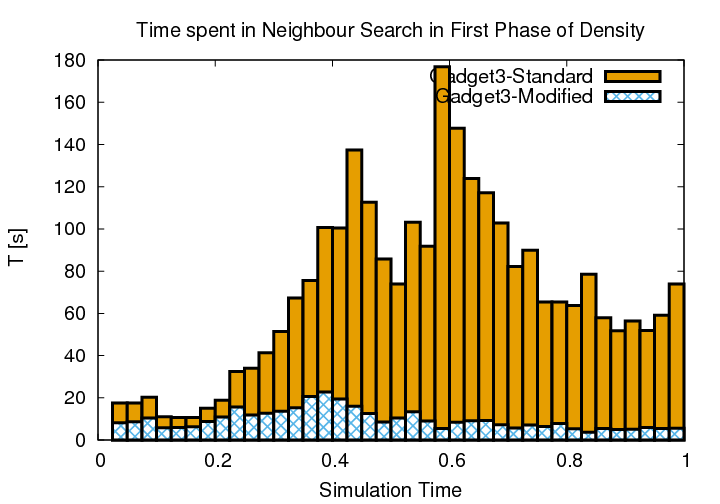}
}
		\caption{Left panel: every bin contains the number of Neighbour Search calls performed in that bin. Right panel: every bin contains the CPU time spent by the Neighbour Search. In both panels the orange (dark) histogram represents the standard version, light blue (light) histogram represents the modified version.}
		\label{fig:ngbs}
\end{figure}

We now investigate quantitatively how the new algorithm affected the performances of the code with respect to the old version. 
To show in details the effect of this new algorithm, we gradually implemented it in various parts of Gadget3 seeing the partial speedups.
First we added the Neighbour Recycling in the First Phase of the Density computations.
Then it has been added on both phases of Density computation, and finally it has been added in both the Hydrodynamics and Density computations.

\subsection{The Test Case} 
We test the algorithm in a cosmological hydrodynamical environment.
We use initial conditions from the \magneticum project \cite{magneticumhp2015}.
To test our algorithm we chosen the simulation \texttt{box5hr}.
This setup has a box size of $18\ Mpc/h$ and  $2\cdot81^3$ particles.
The simulation run on $8$ MPI processes, each with $2$ threads.
The average number of neighbours  is set to $\average{\nngb }=291.$

We have chosen a value of  $f=0.5.$
Using Equation \ref{eq:f}, we obtain $\average{\ncandidates } = 3.375 \average{\nngb}.$
This means that a Tree Walk will now search for $3.375$ more particles compared to the old of Gadget3.
On the other hand such a high theoretical number of particles in a group will definitively justify the overhead of the Neighbour Recycling.
In fact it is inversely proportional to the number of times the tree walk is executed.
Still, such a low ratio between the size of superset of neighbours and the true number of neighbours will not produce a noticeable overhead in the refining of the superset of neighbours.

\subsection{Results} 
\begin{figure}
\centerline{
\includegraphics[width=0.5\textwidth]{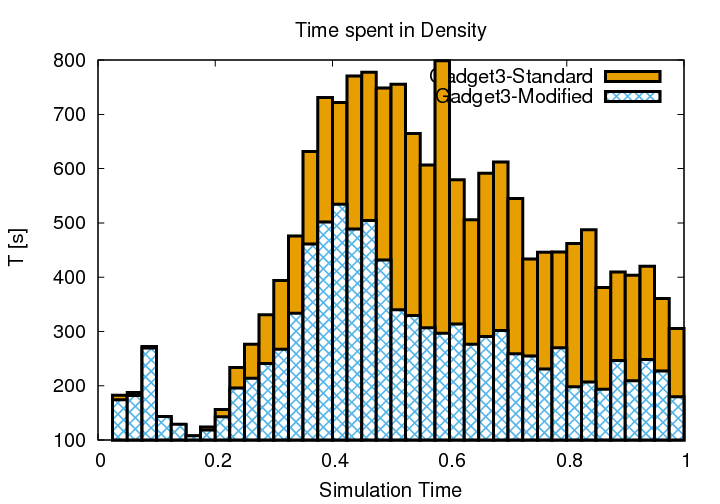}
\includegraphics[width=0.5\textwidth]{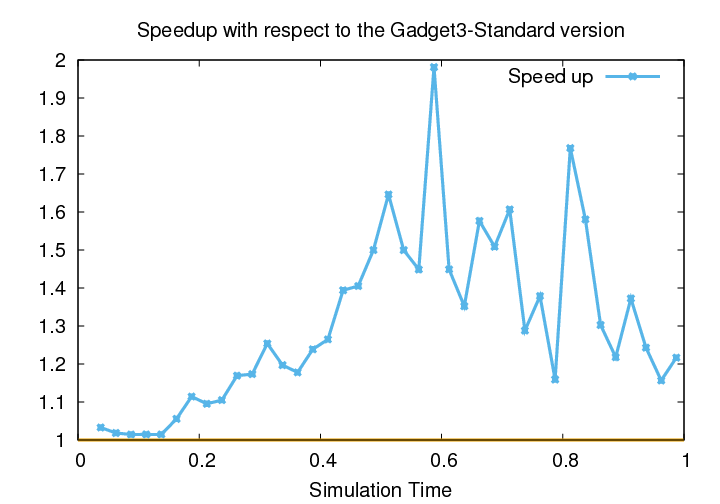}
		}
		\caption{Left panel: every bin contains the time spent in executing the Density module.  The orange (dark)  histogram represents the standard version, light blue (light) histogram represents the modified version. Right panel: speedup of the modified version with respect to the standard version, as a function of the simulation time.}
		\label{fig:frax}
\end{figure}
\begin{figure}
\centerline{
			\includegraphics[width=0.5\textwidth]{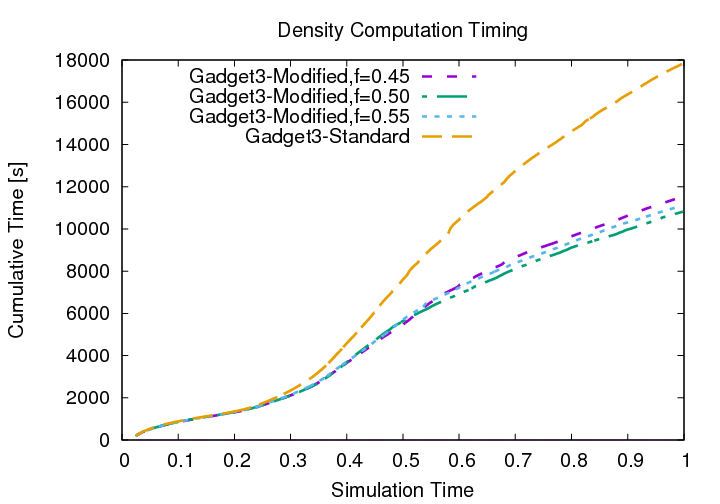}
			\includegraphics[width=0.5\textwidth]{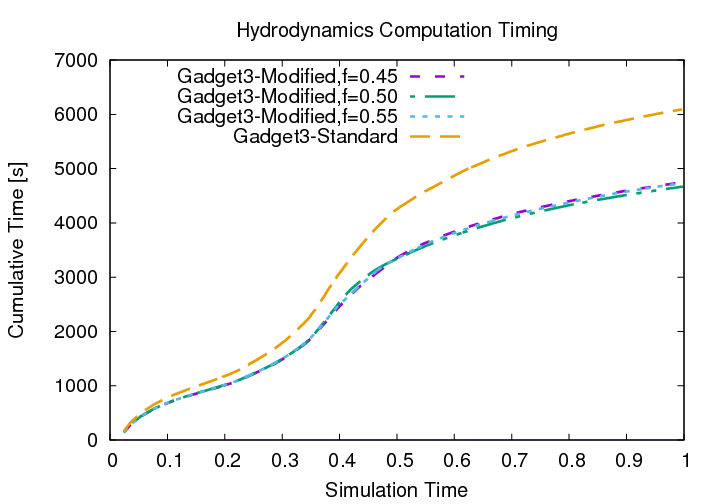}
		}
		\caption{Wall time (in seconds) as a function of the simulation time for different runs: standard version and the new version with $f=0.45,0.50,0.55$. Left panel: Density computation timings. Right panel:  Hydrodynamics computation timings.}
		\label{fig:comparison}
\end{figure}
The algorithm has been first implemented in the first phase of Density module computation of Gadget3.
Figure \ref{fig:ngbs} (left panel) shows the number of Neighbour Search calls performed during the simulation.
The Neighbour Recycling version of the code has roughly the same amount of searches throughout the whole simulation, whereas the old version has a huge peak of Neighbour Search calls around a simulation time of $0.4.$
There, the number of Neighbour Search  calls 
High Performing Code
- Something about communication, and domain decomposition stuff, maybe MPI tricks we used
- different simulations, each many clusters, each many gals
- run analysis on the serverfrom the standard to the modified version,  drops  of a factor of $10.$ 
Compared to the old version, this also means that in this part of the simulation, the average number of particles in a group is $10.$

Theoretically, if all particles within the same sphere were put into the same group, the number of Neighbour Search calls  should drop by a factor of $\average{\ngroup }\simeq230.$ There may be two main reasons why this value is not reached: some time steps do not involve all particles, thus the density of active particles is lower than the density of all particles (which is used to calibrate the radius of the grouping sphere); moreover, the space-filling-curve ordering will leads to particles outside the grouping sphere before the sphere is completely filled.
Those two effects contribute in reducing the number of particles within a grouping sphere, thus increasing the number of Neighbour Search calls.

Figure \ref{fig:ngbs} (right panel) shows the time (in seconds) spent to execute tree walks before and after the modification.
Because the simulation runs on a multi core and using multiple threads, the total time corresponds to the sum of CPU times of all threads.
This plot shows a speedup  that reaches the order of $10$ when the simulation time is approximately $0.4.$
Although the average time of a single Neighbour Search is supposed to be higher, the total time spent for doing the Neighbour Search in the new version is smaller.

The time spent in the density module is shown in Figure  \ref{fig:frax} (left panel).
Here the Neighbour Recycling is  implemented in both the first and the second phases of the density computation.
Unlike previous plots, in this plot the time is the cumulative wall time spent by the code.
As already pointed out, this new version increases the communications between MPI processes. The density module also has very expensive physics computations.
The maximum speedup on the whole density module is larger than a factor of $2.$

\begin{figure}
		\includegraphics[width=\textwidth]{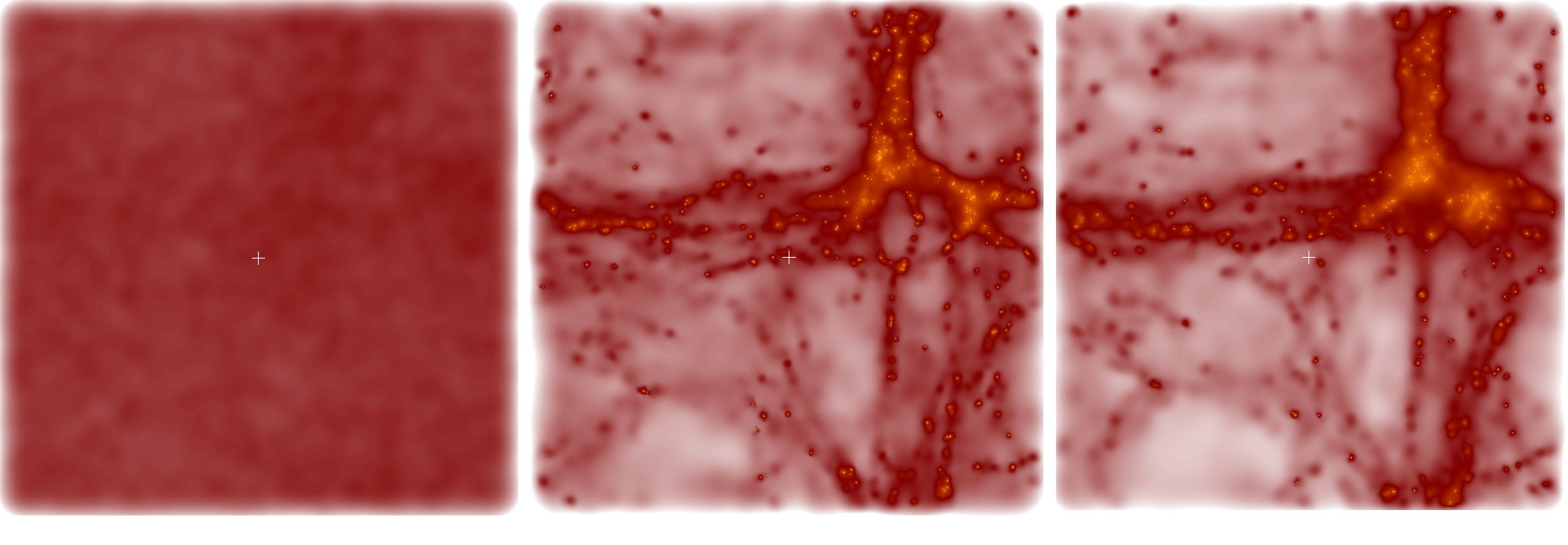}
		\caption{Gas distribution of the cosmological simulation \texttt{box5hr}. Left panel shows the gas distribution of nearly the initial conditions of the simulation; central panel at the middle of the simulation, where clusters start forming; right panel at the end of the simulation. The simulation contains also Dark Matter and stars that have been removed from those plots.}
		\label{fig:snaps}
\end{figure}

Figure \ref{fig:snaps} shows the projected gas distribution in three different phases of the simulation. At the beginning of the simulation gas is distributed homogeneously; this means that the majority of particles are in the same level of the tree.
In the middle panel, voids and clusters can be seen. 
Particles in clusters require smaller time steps, and thus a larger number of Neighbour Search calls.
This is in agreement with the peak of Neighbour Search calls around a simulation time of $0.4$ in Figure \ref{fig:frax}.
This explains why density computations became more intensive for a value of the simulation time greater than $0.4$ (see Figure \ref{fig:comparison}).

Now we check the impact of the Neighbour Recycling on the whole simulation. Figure \ref{fig:frax} (right panel) shows the speedup obtained by implementing the  Neighbour Recycling in both the Density and Hydrodynamics module (the two numerically most expensive modules). The total speedup reaches a peak of $2.2$.

In Figure \ref{fig:comparison} (left panel), using the new approach we see a total cumulative execution time of the Density module of $1.0\cdot10^4s,$ while the standard version has $1.7\cdot10^4s,$ which correspond to a speedup of $1.64.$
Figure \ref{fig:comparison} (right panel) shows the same for the Hydrodynamics module.
The old version spent a cumulative time of $6.0\cdot10^{3}s,$ whereas the new version has $4.6\cdot10^{3}s.$
Leading to a speedup in the hydrodynamics of $1.30.$
The Hydrodynamics module achieved a speedup of $1.30.$
Besides the Density module, a speedup can be seen also at the beginning of the simulation.

Figure \ref{fig:comparison} shows the wall time of the simulation when varying the parameter $f.$
Since we do not knew a priori which value of $f$ will have maximized the speedup, we found it by numerical experiments.
We tried several values of it; values of $f$ near zero gives no speedup, while values of $f$ much greater than one slow down the whole simulation.
In Figure \ref{fig:comparison} there are the timings for the setups with $f=0.45,0.50,0.55.$
The maximum speedup is obtained for $f=0.50$ in both the Density and Hydrodynamics computations.

\section{ Conclusions}

We developed and implemented a way to recycle neighbours to accelerate the Neighbour Search in order to fasten Gadget3.
Our technique should work,  in principle, for any N-Body code with a space-filling-curve ordering of particles. 

This technique groups particles that will be processed one after the other and that are close enough, and makes a single neighbour search for them .
We presented a version of the algorithm that scales the grouping radius with the local density.
This version depends on a constant factor $f.$
We found the value of $f$ that gives the maximum speedup.
In case of the simulation \texttt{box5hr}  of the \magneticum project, corresponds to one half of the searching radius of the single particles.
This radius, of course, depends on the way particles are grouped together.
In this approach we opted for a grouping that depends on the distance from the first particle of the group.
This decision is arbitrary and dictated by the simplicity of the implementation. 

This configuration leads to a speedup of the density computation of $1.64,$ which is known to be one of the most expensive modules in Gadget3.
Implementing this technique in the hydro-force computation too gives a speedup of the whole simulation of $1.34.$

\bibliographystyle{ieeetr} 
\bibliography{ref}

\end{document}